\documentclass{article}

\usepackage{arxiv}

\usepackage[utf8]{inputenc} 
\usepackage[T1]{fontenc}    
\usepackage{hyperref}       
\usepackage{url}            
\usepackage{booktabs}       
\usepackage{amsfonts}       
\usepackage{nicefrac}       
\usepackage{microtype}      
\usepackage{lipsum}
\usepackage{graphicx}
\graphicspath{ {./images/} }

\title{Automated Multiclass Cardiac Volume Segmentation and Model Generation}

\author{
  Erik Gaasedelen \\
  Department of Surgery\\
  University of Minnesota\\
  Minneapolis, Minnesota \\
  \texttt{gaas0012@umn.edu} \\
   \And
 Alex J. Deakyne \\
  Department of Surgery\\
  University of Minnesota\\
  Minneapolis, Minnesota \\
  \texttt{deaky007@umn.edu} \\
     \And
 Paul A. Iaizzo \\
  Department of Surgery\\
  University of Minnesota\\
  Minneapolis, Minnesota \\
  \texttt{iaizz001@umn.edu } \\
}

\begin{document}
\maketitle

\begin{abstract}
Many strides have been made in semantic segmentation of multiple classes within an image. This has been largely due to advancements in deep learning and convolutional neural networks (CNNs). Features within a CNN are automatically learned during training, which allows for the abstraction of semantic information within the images. These deep learning models are powerful enough to handle the segmentation of multiple classes without the need for multiple networks. Despite these advancements, few attempts have been made to automatically segment multiple anatomical features within medical imaging datasets obtained from CT or MRI scans. This offers a unique challenge because of the three dimensional nature of medical imaging data. In order to alleviate the 3D modality problem, we propose a multi-axis ensemble method, applied to a dataset of 4-cardiac-chamber segmented CT scans. Inspired by the typical three-axis view used by humans, this technique aims to maximize the 3D spatial information afforded to the model, while remaining efficient for consumer grade inference hardware. Multi-axis ensembling along with pragmatic voxel preprocessing have shown in our experiments to greatly increase the mean intersection over union of our predictions over the complete DICOM dataset.
\end{abstract}


\section{Introduction}
\label{sec:introduction}
\subsection{Anatomical Segmentation}
Medical Imaging datasets, such as DICOM image datasets produced from CT or MRI scans, contain a wealth of information about a patient and their anatomy. This information is invaluable to medical device companies and physicians alike. However, this information, especially when dealing with complex anatomical features, is not easily digested. DICOM datasets are two dimensional images that are stacked to create a three dimensional representation of the patient anatomy. Analyzing a DICOM dataset is traditionally done by sweeping through these two dimensional slices in order to get an idea of the three dimensional anatomy. This has lead to the development of tools that can be used to segment anatomical features of interest within DICOM scans. All three axis of the scan can be viewed simultaneously to provide the user with all the spatial information in the scan while segmenting. Segmentations not only highlight the anatomical feature, but certain software packages allow for the creation of 3D models from the 2D segmentations. An example of such a pipeline is depicted in Figure 1 below. These models have wide use in both pre-surgical planning and medical device design. They offer a clear representation of the anatomical feature while retaining all the spatial information in the scan. These models can be easily used to generate anatomical measurements and generate statistics about a certain patient population.

Although anatomical segmentations are becoming an invaluable information source in the medical field, they are incredibly difficult to produce. Interpretation of simple anatomical features within a DICOM dataset takes time and practice. Often, there are artifacts present in the scans and the distinction between anatomical features can be nebulous. It takes an expert in the relevant anatomies to produce quality segmentations. Also, there are few anatomical segmentation tools that work well, and their availability is gated behind incredibly expensive licensing fees. These restrictions have granted few people with the necessary skill set to accurately interpret DICOM scans and produce quality anatomical models. 

Unfortunately, anatomical segmentation is a time consuming process, even for highly skilled individuals. This creates a prohibitive bottleneck in many workflows that wish to utilize anatomical segmentations and 3D models. For surgical planning, a physician must receive a segmented model of the patient anatomy between the pre-procedure scan and the start of the given procedure. To date, this is often logistically impossible which in turn could be considered to restrict the quality of care available to the patient. Further, in the medical device industry, the anatomy of a patient population must be well understood in order to make informed decisions about how to design a new device. This often requires the production of hundreds of anatomical segmentations, which are then used to make measurements to model the statistical information of the patient population. This can be such an arduous process that many companies are forced to make their decisions on fewer segmentations than they would like.

\begin{figure}[ht]
  \centering
  \includegraphics[width=8cm, height=6cm]{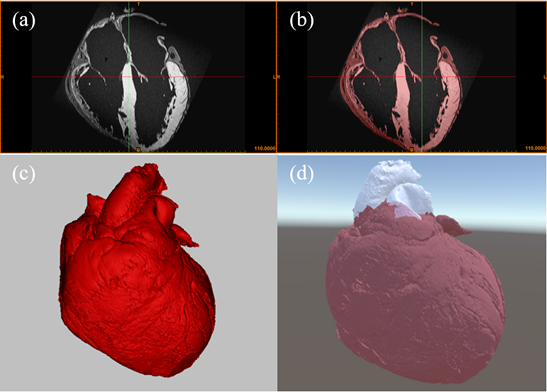}
  \caption{ Flowchart diagram of proposed mixed reality methodology. (a) A DICOM scan in MIMICS depicting a cross sectional view of a heart. (b) 2D mask of the heart tissue generated through thresholding functions. (c) The resulting 3D model of the heart created by the 2D masks which can then be 3D printed. (d) The same 3D model of the heart is imported into Unity3D and a virtual reality environment of this given heart is generated.}
  \label{fig:model pipeline}
\end{figure}

\subsection{Deep Learning}
Deep learning has established itself as the current state of the art for supervised image segmentation. The application emerged by combining traditional classification networks with a deconvolutional layer set. While initially ineffective, the addition of skip connection between downsampling and upsampling passes in an architecture called U-Net significantly improved upon the conv-deconv model, producing state of the art results on diverse datasets [1]. The skip-connection discovery closely parallels the development of residual connections in traditional classification networks. As demonstrated in research on resnet and densenet architectures, these strategies not only allow for deeper networks without vanishing gradients, but significantly speed up model convergence and improve model accuracy [2,3].

\begin{figure}[ht]
    \centering
    \includegraphics[width=8cm]{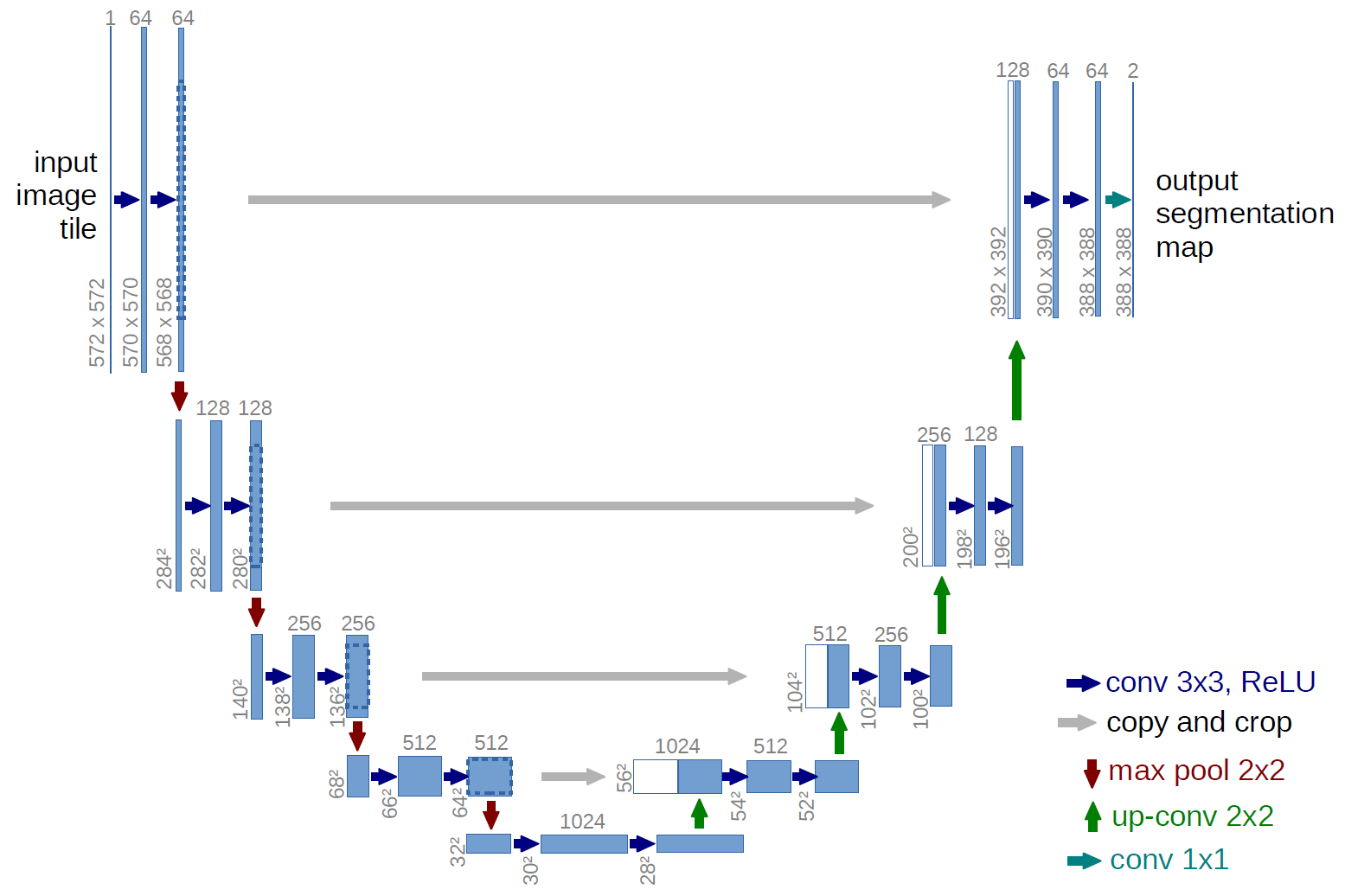}
    \caption{U-Net Architecture}
    \label{fig:unet}
\end{figure}

\begin{figure}[ht]
    \centering
    \includegraphics[width=8cm]{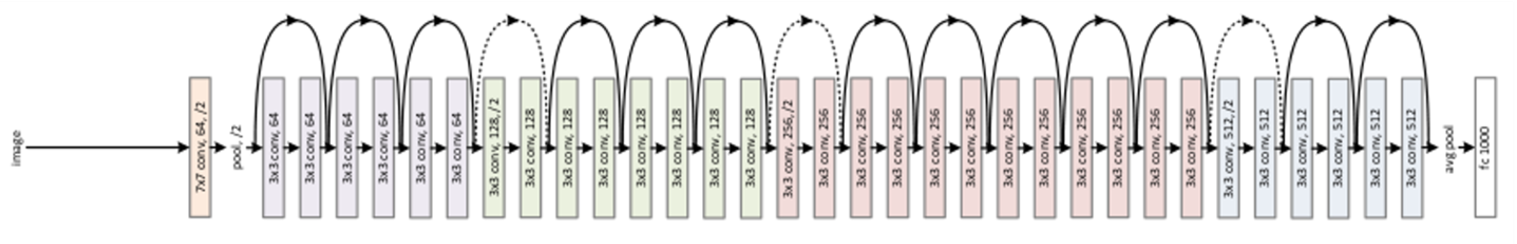}
    \caption{ResNet-34 Architecture}
    \label{fig:resnet}
\end{figure}

For binary segmentation, the standard U-Net architecture works well for most domains of interest. For multi-class segmentation, state of the art models resort to some additional modifications. These models are usually variants of the feature pyramid network architecture [4]. The benefit of feature pyramid networks usually becomes apparent as the number of classes increases. Unfortunately, feature pyramids require more GPU memory, are slower to train, and typically produce outputs that can be significantly downsampled compared to the original image. It is unclear why U-Nets and feature pyramid networks have such a performance discrepancy in the multi-class domain.

\subsection{Automatic Segmentation Using Deep Learning}
Medical imaging problems commonly require that one extends the 2D segmentation problem to a 3D space. Considering just the desired output of a 3D segmentation problem, it is natural to consider the utilization of a 3D U-Net by simply replacing two-dimensional layers with corresponding three-dimensional layers [5]. The downsides of this strategy is that it is prohibitively expensive for GPU memory and limits the usability of transfer learning for lack of pretrained models on 3D datasets.

Performing a multi-class task on modest hardware resources is not without hope, however. For manual segmentation, software packages provide views of the axial, coronal, and sagittal axes. This provides users with enough information to create good segmentation masks, even though it may take significant time to cross check all planes and neighboring depths.

\begin{figure}[hb]
    \centering
    \includegraphics[width=8cm]{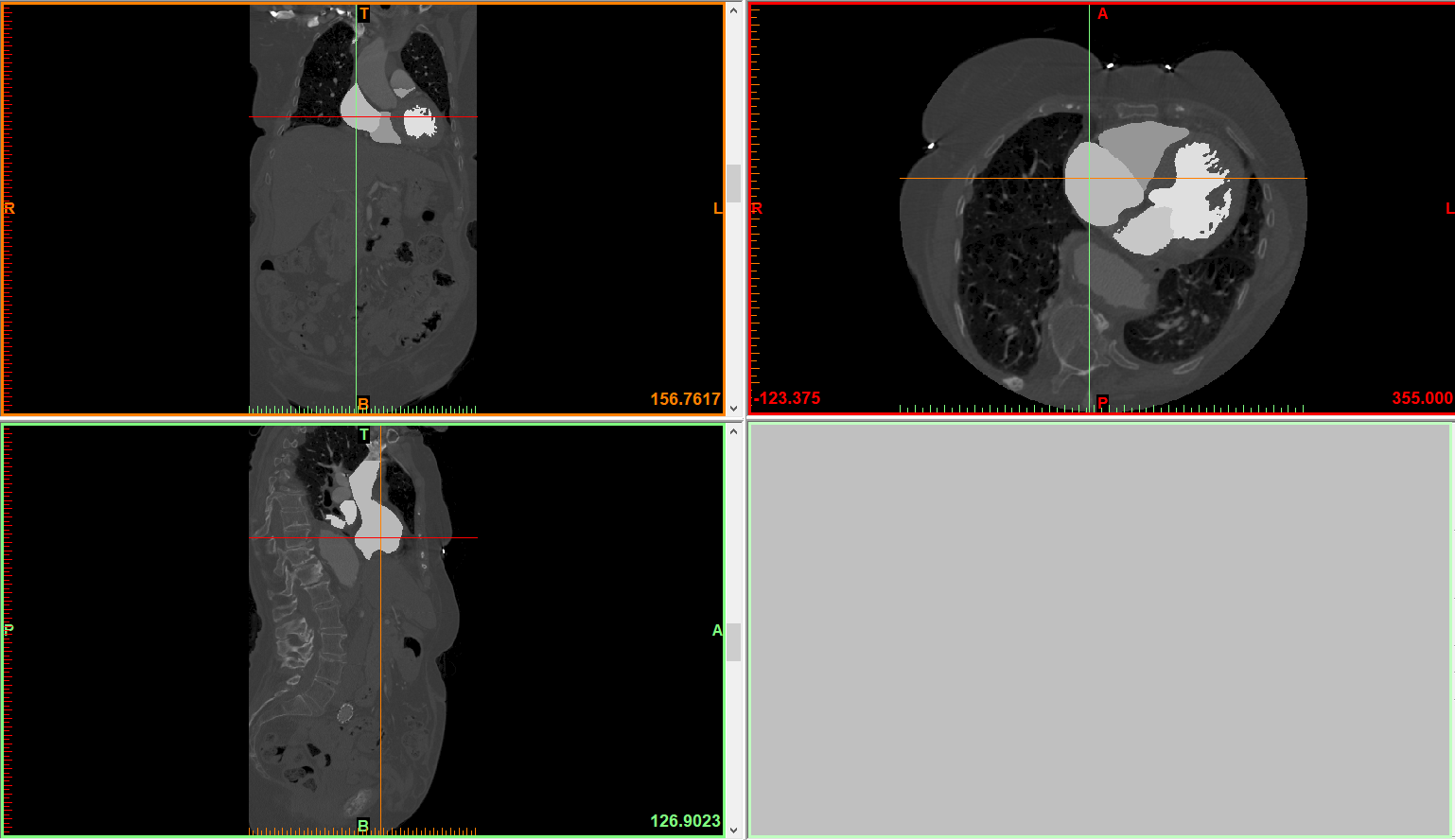}
    \caption{Screenshot of manual segmentation software, Mimics, developed by Materialise}
    \label{fig:mimics}
\end{figure}

The dataset we analyzed here included 46 contrast-enhanced cardiac CT scans with each of the four chambers of the heart segmented. Normally 3D medical imaging datasets comprise of MRI scans which usually have much better resolution and are safer to acquire. While MRI uses a magnetic field to construct an image through hydrogen spin resonance, CT technology uses x-ray radiation to determine at what depths radiation permeates and scatters. Since bones and radio-opaque contrast scatter x-rays more, their pixel values are brighter in the resulting CT image.

\begin{figure}[ht]
    \centering
    \includegraphics[width=8cm]{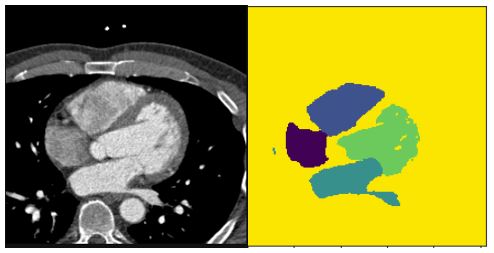}
    \caption{Dataset example of a scan slice (left) and corresponding mask (right)}
    \label{fig:data}
\end{figure}

CT scans were then recorded on the Hounsfield scale, represented by an integer between -1000 and 30000. For pretrained neural networks, images were usually normalized after converting 8 bit (0 through 255 valued) images to be between zero and one. Pixel values between -1000 and 3000 have a much larger range and therefore must be addressed to make easily compatible with pretrained models [6].

Here we attempted to address all of these difficult problems while retaining the speed, efficiencies, and pretrained weights of a traditional U-Net construction. We found that through simple domain guidance, pragmatic dataset construction, and axis averaging, we achieved highly accurate results, with the abilities to correctly segment and label minute details of cardiac anatomies.

\section{Methods}
\label{sec:method}
\subsection{DICOM Data Preprocessing}
First we must address the problem of the large spread of the Hounsfield scale. The purpose of consistent data normalization in a transfer learning scenario is not necessarily to match color distributions of the previously trained imageset, but rather to allow color perception to be consistent between datasets. This analogy becomes difficult in the case of a CT scan where its brightness units are most analogous to grayscale color values. Fortunately, we do not need the complete spread of Hounsfield units; only a small range of brightness values that would be most important for distinguishing the volumes of interest. We therefore clipped the Hounsfield units between -200 and 500, before allowing that range to take the values between 0 and 1 prior to normalization. These values roughly correspond to airy tissue like lung (-200)  up to cortical bone (500), which places contrast dye nicely between these values. Such a step is very important as it counteracts under or over saturation in the images which would be difficult for a pretrained network to overcome.

For ease of processing we also converted all DICOMs and masks to PNG files. Masks were produced simply by assigning the pixel values in each slice to 0, 1, 2, 3, and 4 corresponding to the right atrium, right ventricle, left atrium, left ventricle, and background respectively.

\subsection{Architecture}
For our architecture we choose a U-Net. The U-Net incorporates a pretrained resnet-34 network as the encoder, as implemented in the fastai library. [9] We choose resnet-34 in order to maximize our batch size on a single Nvidia 1080ti. Although, with more computing resources, a resnet-50 or feature pyramid network could have an added benefit. These considerations are left for the discussion.

\subsection{Training}
For training we used the one-cycle policy which is included in the fastai library [7, 9]. We first trained the network with frozen weights for 10 epochs. We then lowered the learning rate and continued training for 10 epochs with unfrozen weights. For image augmentations we included ten degree rotations, zooms, lighting changes, and random crops at 256x256. To do this we sampled from our dataset on all axes including axial, coronal, and sagittal slices. We used cross entropy loss for our optimization objective.

\begin{figure}[ht]
    \centering
    \includegraphics[width=8cm]{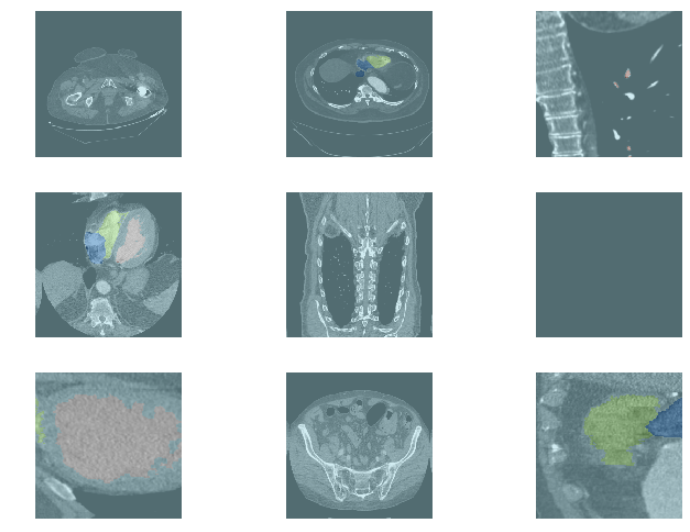}
    \caption{Example Augmented Training data}
    \label{fig:train data}
\end{figure}

\subsection{Evaluation}
For each study we evaluated each slice on all axes. Before running them through the network, however, we enforced a maximum dimension of 256x256. This was followed by a slight resize, if necessary, to ensure each dimension was divisible by 32 to ensure proper divisibility of shape throughout the network. Once all slices from each axis were evaluated, the resulting activations were averaged between all planes. This output was then used to determine the label for every voxel in the 3D tensor, and then subsequently resized back to its original shape before computing metrics.

\subsection{3D Model Generation}
We used the marching cubes implementation in the scikit-image library to create a surface mesh for each chamber of the heart. These meshes were then used to 3D print select examples. The DICOM metainformation for pixel spacing and distance between slices was also used to output meshes that were of consistent sizing to the patient’s native anatomy.

\begin{figure}[hb!]
    \centering
    \includegraphics[width=6cm, height=6cm]{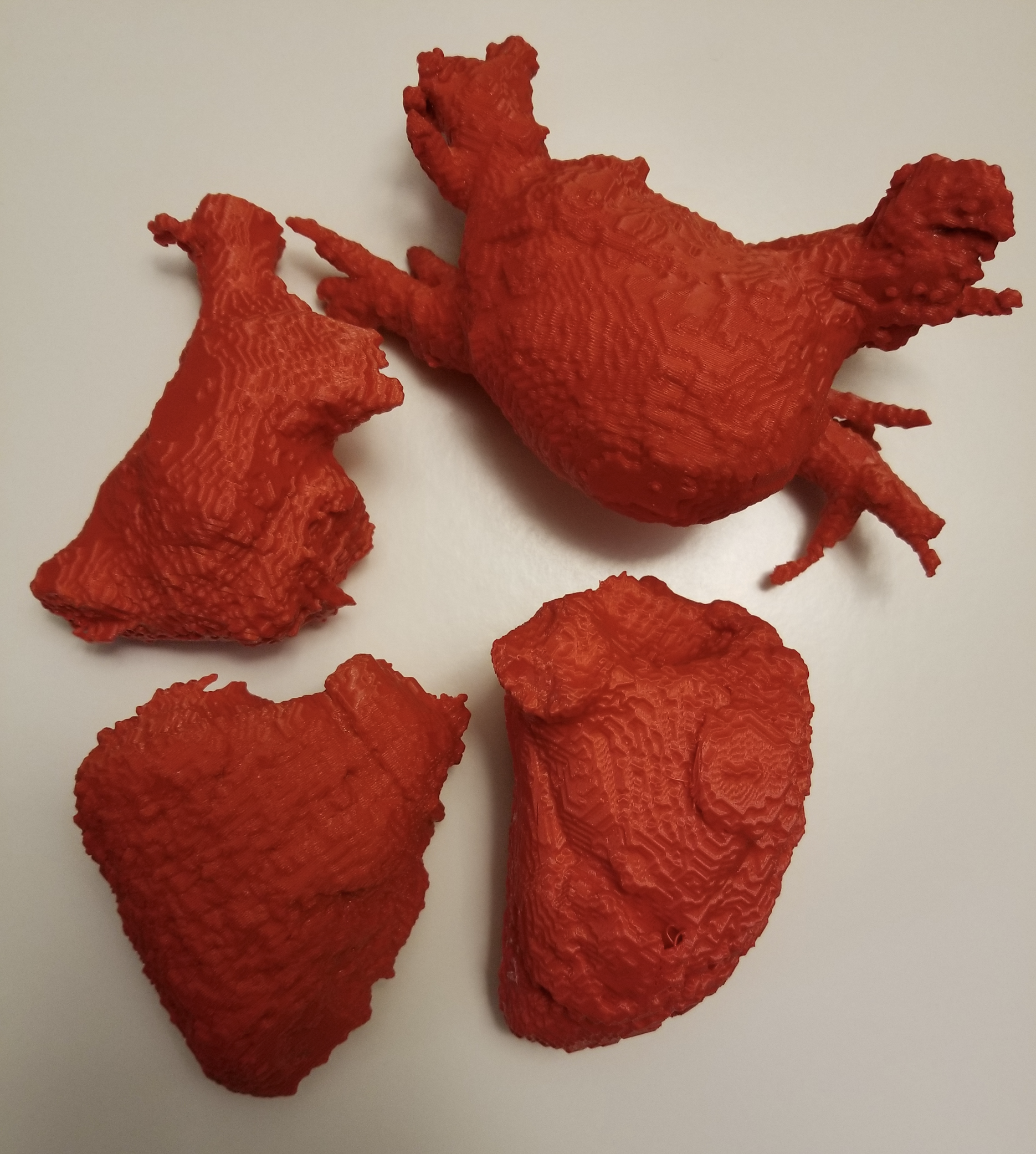}
    \caption{3D Model of network segmented heart. Right atrium (upper left), left atrium (upper right), right ventricle (lower left), and left ventricle (lower right).}
    \label{fig:3d print}
\end{figure}

\section{Results}
Our primary method for achieving highly accurate results was through our multi-axis ensembling approach. When we allowed each axis to average into the final output we achieved a mean intersection over union of 0.861. To compare performance without such averaging, we also calculated the same metric between axial, coronal, and sagittal exclusively. None of the axes performed as well in isolation with the axial axis achieving the next best score with a 0.851 mIOU.

\begin{table}
 \caption{ Comparison of mIOU for each chamber, axis, and multi-axis ensemble}
  \centering
  \begin{tabular}{*{6}{c}}
    \toprule
    & Right Atrium & Right Ventricle & Left Atrium & Left Ventricle & Full Heart \\
    \midrule
    \textbf{Ensemble} & \textbf{0.8872} & \textbf{0.8257} & \textbf{0.9109} & \textbf{0.8803} & \textbf{0.8610}     \\
    Axial & 0.8376 & 0.8122 &  0.8798 & 0.8736 & 0.8508     \\
    Coronal & 0.8036 & 0.7786 & 0.8366 & 0.8598 & 0.8196  \\
    Sagittal & 0.72228 & 0.7846 & 0.8351 & 0.8620 & 0.8011 \\
    \bottomrule
  \end{tabular}
  \label{tab:table}
\end{table}

Each chamber performed with different average mIOUs throughout our validation dataset. Although ensemble consistently performed better than the individual axis.  The ventricular chambers typically performed better than the atrial chambers. This is probably because the ventricles are better defined when it comes to where its boundaries are delineated. For atria, decisions must be made as to where they end with respect to venous inlets. The left atrium, for instance, is connected to many branching pulmonary veins. For a human segmenter, it can be ambiguous as to where these veins should no longer be segmented.

3D model inspection, however, is the most clear when it comes to showing the minute details our model is able to pick out of the DICOMs. The 3D models we generated were able to clearly show fine details of such structures of the left atrial appendage, semilunar valves, and the extensive trabeculae and papillary muscles that occupy the free wall of the left ventricle. 3D models from the same scan were generated utilizing just the axial slices and using the multi-axis ensembling technique. These models, along with the ground truth model, are displayed below. 

\begin{figure}[hb!]
    \centering
    \includegraphics[width=10cm, height=4cm]{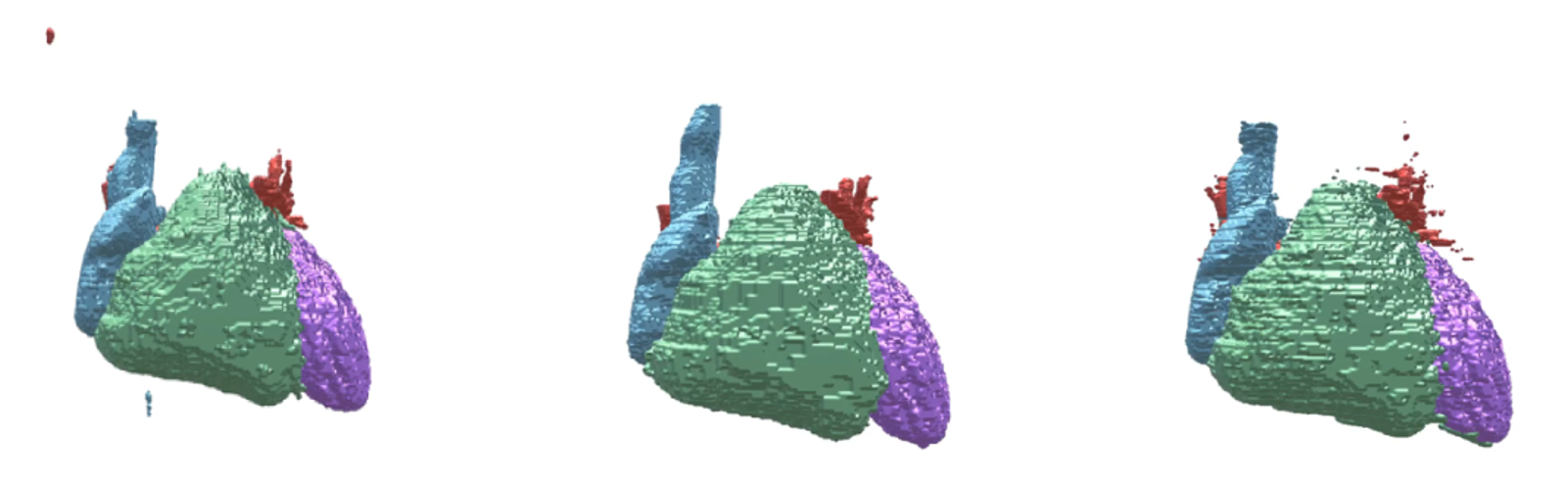}
    \includegraphics[width=10cm, height=4cm]{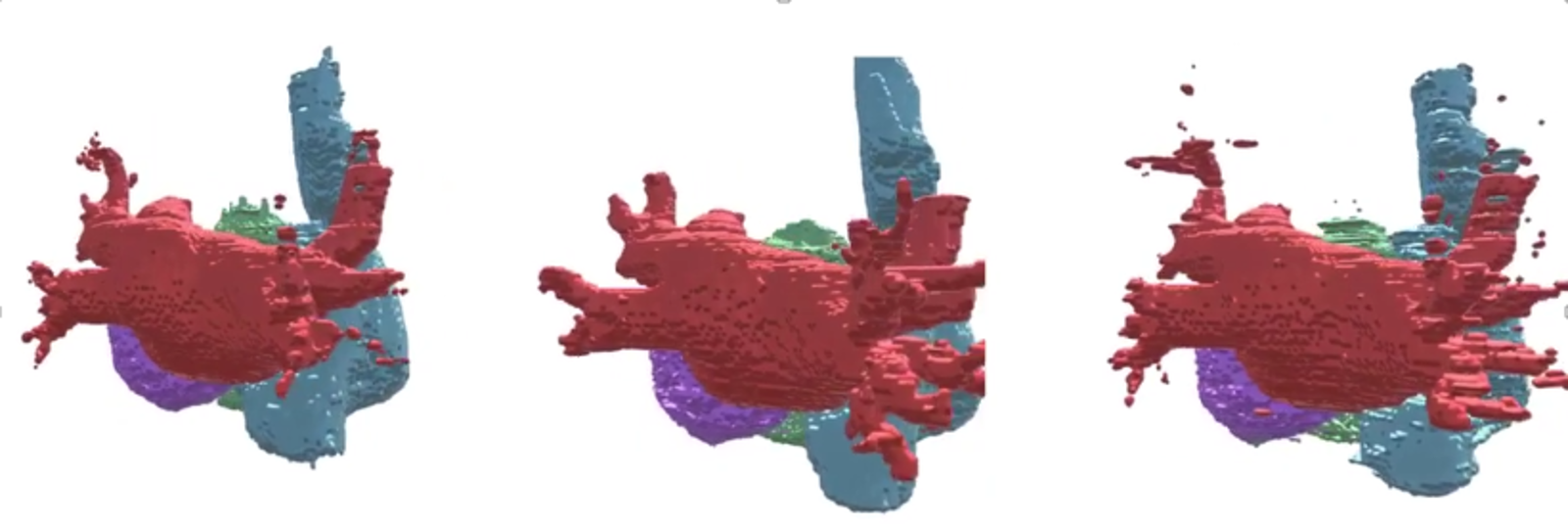}
    \caption{Anterior (top) and posterior (bottom) views of multi-axis ensemble (left), ground truth (middle), and axial (right) 3D models generated from the same patient scan.}
    \label{fig:4heart 3d models}
\end{figure}

\section{Conclusion}
Creating an unambiguous metric for a task like anatomical segmentation can be difficult. Currently, a human manual segmenter needs to make many decisions as to where the boundaries of a chamber are. For example, superior and inferior vena cava can be inconsistently segmented as well as the pulmonary veins. It can also be difficult to decide when to segment over what looks like muscular bands within the heart, especially when image quality varies between scans. Therefore using human segments as a substrate for objective minimization and metric evaluation can be tricky. Nevertheless, neural networks are robust to errors and variation that are contained within datasets. Although we achieved results that matched very well with human generated masks, ultimately our performance must be taken into context of visual inspections and whatever applications are appropriate: e.g., for anatomical education these are highly useful today, but for pre-clinical planning further evaluation is needed.

It should be noted, that a major consideration that went into our model design was the resource limitations of our computer. A U-Net with a resnet-34 backbone was more than sufficient for creating highly accurate results, but there are some opportunities for improvement. Larger encoder networks such as resnet-50, resnet-101, or squeeze excitation networks could give a boost to the performance of our model [8].

Feature pyramid networks are also promising substitutes for U-Nets. In our problem description we only needed to care about five unique classes. For anatomical datasets with ten or more classes, we suspect feature pyramid networks would begin to see benefits that justify the acquisition of more compute power.

Cardiac segmentation is a great example where macroscopic and fine details must be integrated together to get a proper understanding of the underlying anatomies. While the left ventricle is large, its internal structure is composed of delicate muscle bands which can be an important factor for medical device design and surgical procedures. A further challenge will be datasets where classes are imbalanced between very large structures and very small structures. This is where tricks like feature pyramid networks and class specific sampling frequency will be especially important.

The applications for scalable and fast anatomical segmentation are vast. Presurgical planning and human centered medical device design just scratch these surfaces. Opportunities for large simulation datasets, live surgical visualizations, statistical shape modeling, and mesh-based automated diagnosis all become approachable when such systems become readily available. Important next steps will be to improve our model performance with more computational resources, expand our dataset to additional organs of interest, investigate opportunities for rapid visualization and analytical investigations of generated models, and develop additional tools to detect or classify different anatomical regions of interest in a more diagnostic mode.

\bibliographystyle{unsrt}  
\nocite{*}
\bibliography{references}








\end{document}